# Social and behavioral determinants of health in the era of artificial intelligence with electronic health records: A scoping review

Anusha Bompelli[#], Department of Pharmaceutical Care & Health Systems, University of Minnesota, bompe001@umn.edu

Yanshan Wang[#], Department of Health Information Management, University of Pittsburgh, yanshan.wang@pitt.edu

Ruyuan Wan, Department of Computer Science, University of Minnesota, wanxx199@umn.edu

Esha Singh, Department of Computer Science, University of Minnesota, sing0640@umn.edu

Yuqi Zhou, Institute for Health Informatics and College of Pharmacy, University of Minnesota, zhou0357@umn.edu

Lin Xu, Carlson School of Business, University of Minnesota, xu000532@umn.edu

David Oniani, Department of Computer Science and Mathematics, Luther College, onianidavid@gmail.com

Bhavani Singh Agnikula Kshatriya, Department of Health Sciences Research, Mayo Clinic, AgnikulaKshatriya.BhavaniSingh@mayo.edu

Joyce (Joy) E. Balls-Berry, Department of Neurology, Washington University St. Louis, j.balls-berry@wustl.edu

Rui Zhang*, Institute for Health Informatics, Department of Pharmaceutical Care & Health Systems, University of Minnesota  zhan1386@umn.edu

# Equally contributed

*Corresponding author

**Abstract**

**Background:** There is growing evidence that social and behavioral determinants of health (SBDH) play a substantial effect in a wide range of health outcomes. Electronic health records (EHRs) have been widely employed to conduct observational studies in the age of artificial intelligence (AI). However, there has been little research into how to make the most of SBDH information from EHRs.
**Methods:** A systematic search was conducted in six databases to find relevant peer-reviewed publications that had recently been published. Relevance was determined by screening and evaluating the articles. Based on selected relevant studies, a methodological analysis of AI algorithms leveraging SBDH information in EHR data was provided.
**Results:** Our synthesis was driven by an analysis of SBDH categories, the relationship between SBDH and healthcare-related statuses, and several NLP approaches for extracting SDOH from clinical literature.
**Discussion:** The associations between SBDH and health outcomes are complicated and diverse; several pathways may be involved. Using Natural Language Processing (NLP) technology to support the extraction of SBDH and other clinical ideas simplifies the identification and extraction of essential concepts from clinical data, efficiently unlocks unstructured data, and aids in the resolution of unstructured data-related issues.
**Conclusion:** Despite known associations between SBDH and disease, SBDH factors are rarely investigated as interventions to improve patient outcomes. Gaining knowledge about SBDH and how SBDH data can be collected from EHRs using NLP approaches and predictive models improves the chances of influencing health policy change for patient wellness, and ultimately promoting health and health equity.

## 1. Introduction

Social determinants of health are conditions in the environments in which people are born, live, learn, work, play, worship and age that affect a wide range of health, functioning and quality of life outcomes and risks[1]. The social determinants of health are categorized into five key categories: economic stability; education access and quality; social and community context; neighborhood and built environment; and healthcare access and quality[1]. As our population is becoming more diverse, there is growing evidence demonstrating the significant impact of SDOH on various healthcare outcomes such as mortality[2], [3], morbidity[4], life expectancy[3], healthcare expenditures[5], health status, and functional limitations[6]. For example, a study showed that SDOH factors, including education, racial inequality, social support, and poverty, accounted for more than a third of the estimated annual deaths in the United States[6], [7].

It is not only necessary to overcome social and behavioral determinants of health (SBDH) in order to enhance public health, but also to eliminate health inequalities that are often entrenched in social and economic inequalities. One way to address this is by integrating SBDH into the electronic health record (EHR). Increased use of EHR systems in healthcare organizations has facilitated secondary use of EHR data through artificial intelligence (AI) techniques to improve patient care outcomes[8], via clinical decision support systems, chronic disease management, and patient education. Most recently, AI methods were used to propose candidate drugs for COVID-19[9].

SBDH information in the EHR is stored in both structured (e.g., education, salary level) and unstructured formats (e.g., social history in clinical notes). Since there is no standardized framework for recording SBDH information and such information is usually incompletely recorded[10] in a structured format, it is often difficult to identify SBDH present in an unstructured format and to establish a connection between SBDH and disease or health outcomes. Approaches that leverage natural language processing (NLP) tools to extract SBDH information stored in an EHR in an unstructured format are still limited. Prior literature reviews on SBDH mainly focused on integration of SBDH into EHR[11], and impact of SBDH in risk prediction, the role of SBDH in mental health[8], availability and characteristics of SBDH in EHR[12]. None of these reviews discussed the AI methods using SBDH in EHR data. This paper provides a systematic literature review of the SBDH factors, the relationship between SBDH and disease, and the NLP techniques used to extract SBDH information using EHR data.

## 2. Methods

### 2.1 Data sources and search strategies

This systematic literature review followed the Preferred Reporting Items for Systematic Reviews and Meta-Analyses (PRISMA) guidelines. A comprehensive search of several databases from 2010 to October 15th, 2020, English language, was conducted. The databases included Ovid MEDLINE(R) and Epub Ahead of Print, In-Process & Other Non-Indexed Citations, and Daily, Ovid EMBASE, Scopus, Web of Science, the ACM Digital Library, and IEEE Xplore. The editorial, erratum, letter, note, comment article types were excluded. The search patterns used in these databases were consistent. We iteratively updated our searching keywords by searching for relevant articles and identifying specific SBDH-related keywords from those articles and repeating the process. The final keywords used in the search query are shown in **Table 1**, and the query was implemented by an experienced librarian. A detailed description of the search strategies used is provided in the Supplemental Table 1.

**Table 1.** Searching strategies used to retrieve the literature.

| Category | Keywords |
|---|---|
| EHR terms | electronic health record OR electronic health records OR electronic medical records OR electronic medical record OR EHR OR EMR |
| SBDH terms | socioeconomic health or (avail* adj3 care) or (avail* adj3 healthcare) or (health* adj3 access*) or (unisur* adj3 health) or "access care" or "access healthcare" or crowding or "determinants of health" or diet or education or "education achieve*" or "education status" or employment or "environmental factors" or "financial difficult*" or "financial problem*" or "food insecurity" or "health literacy" or homeless or homelessness or housing or "housing instability" or Incarceration or "income difference" or indigent or "insurance health" or "insurance status" or "job insecurity" or jobless or "lack of educational attainment" or lifestyle or "low income" or marginalized or nutrition or "occupational status" or overcrowding or "physical activity" or poverty or "psychosocial depriv*" or "public safety" or "racial discrimination" or racism or "rural health" or SDH or SDOH or SES or "social and behavioral determinants of health" or "social behavior" or "social depriv*" or "social determinants" or "social determinants of behavior" or "social determinants of health" or "social disadvantage" or "social disparity" or "social economics" or "social environment" or "social exclu*" or "social factor*" or "social |
| | gradient*" or "social position" or "social support" or "social variation" or socialeconomics or socioeconomic or "socioeconomic circumst*" or "socioeconomic factor*" or "socioeconomic gradient*" or "socioeconomic position" or "socioeconomic status" or "socioeconomic status socioeconomic variable" or "standard living" or transportation or "underinsure* health" or underprivilege* or unemployed or unemployment or "vulnerable communit*" or "vulnerable group*" or "vulnerable people" or "vulnerable person*" or "vulnerable population* |
| AI terms | NLP OR natural language processing OR information extraction OR named entity extraction OR named entity recognition OR co-reference resolution OR relation extraction OR text mining OR artificial intelligence OR machine learning OR deep learning OR predictive modeling OR AI |

## 2.2 Article selection

After obtaining potential articles, the following steps were abstract screening and full text screening. Article exclusion criteria included: duplicates; conference abstracts; unavailable full text; and articles unrelated to SBDH, AI, or EHR. Two reviewers used the eligibility criteria to screen the articles for selection from title and abstract screening in the first round. In the second round, seven reviewers reviewed the papers for full-text and relevance, with the workload evenly distributed among them. The selected studies from the second round were reviewed further by two reviewers. All conflicts that occurred during the screening rounds were discussed until consensus was reached.

## 2.3 Data extraction from included studies

Two reviewers (RZ and YW) developed and specified the data elements to be retrieved, while the remaining seven reviewers extracted the data from the full-text articles. The following data items were captured in a Supplementary Table 2 entitled, "Raw File: Social and behavioral determinants of health in the era of artificial intelligence with electronic health records: A scoping review": citation, country, SBDH data source, clinical note type, disease, disease category (ICD 10), SBDH focus, SBDH category, SBDH level, SBDH role, study cohort (patients), study cohort (clinical sites) and AI methods.

## 2.4 Data synthesis and analysis

SBDH research is an interdisciplinary field that combines healthcare, social science, and informatics. Based on the articles identified, our data synthesis was motivated by an approach to gain insight into how SDoH affect disease risk or onset prediction using AI and review how effective current NLP systems are at extracting various types of SBDH We began by examining the general characteristics of the 79 included studies, such as publication trend and journal venues. We examined the coverage of SDOH and disease categories and illustrated their relationships to provide insight into how SBDH characteristics influence healthcare-related statuses. Furthermore, we investigated NLP approaches for extracting SDOH and predictive models for predicting outcomes using SDOH characteristics. Supplementary Table 1 contains further information about the included articles and the review analysis. Regarding software: Zotero was utilized for citation management, Microsoft Excel for data extraction and collection, and RAWGraphs for data analysis.

3. **Results**

## 3.1 Identification of included studies

A total of 1,433 articles were retrieved from five libraries, of which 643 articles were found to be unique. The articles were then filtered manually based on the title and abstract to check whether articles are related to SDOH and AI based on the EHR data. 283 articles remained for subsequent full-text reviews. The inclusion criteria for the target publications are: 1) AI techniques are used, 2) the SDOH information is from the EHR data, and 3) EHR data is in English. Articles without full text were excluded. After this full-text screening process, 79 articles were selected in this scoping review. A flow chart of the article selection process is shown in **Figure 1**.

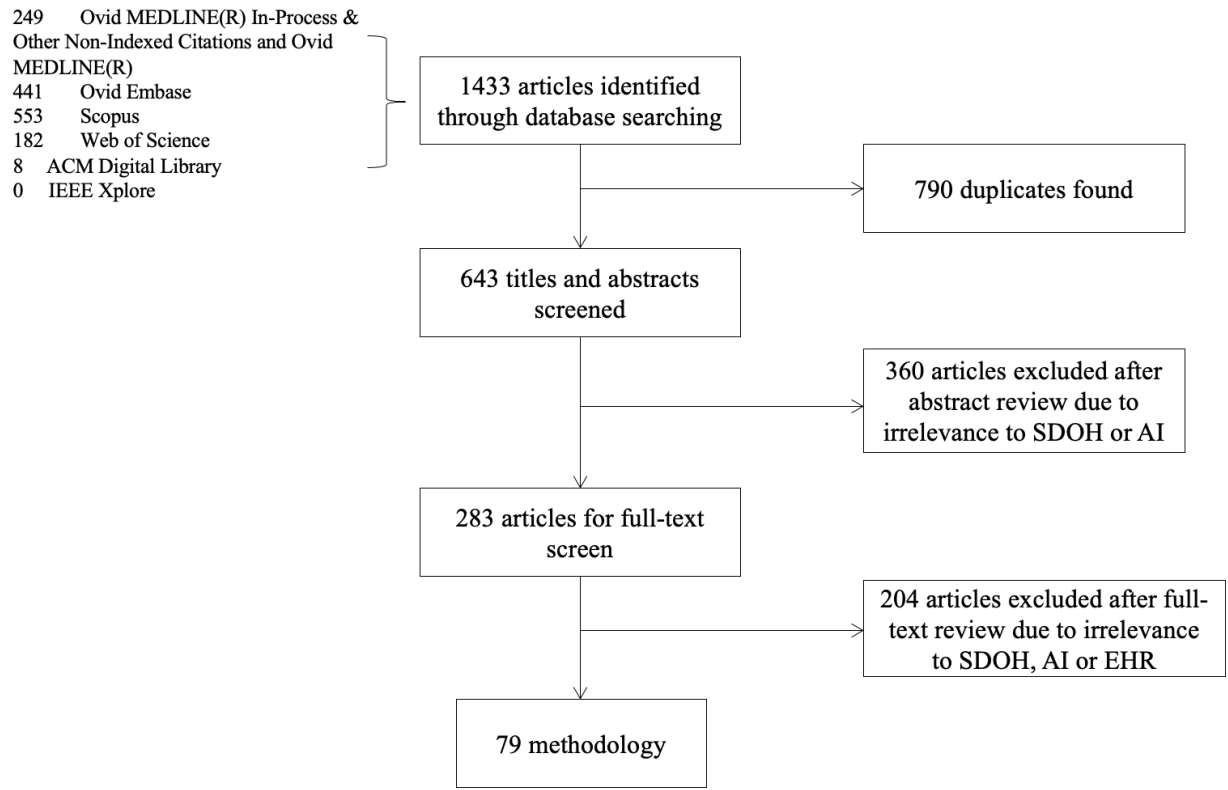

**Figure 1.** The flowchart of the article selection process.

## 3.2 General characteristics of the included studies

The general characteristics of the included studies are summarized in **Table 1**. All the included studies are published between 2012 and 2021 (**Figure 2**) with an increasing number of publications from 2012 to 2020. The current review includes publications across 11 countries (**Figure 3**), with most of the contribution from the United States (77%). Investigation of the publication venues indicates the research communities that utilize SDOH from EHRs by leveraging informatics techniques. The type of venues for conferences are determined through manually researching conference information. The venues for journals are determined through Scimago and Clarivate Analytics. All studies are generally divided into four different types: 1) Clinical (n=31); 2) Informatics (n=33); 3) Social Science (n=12) and 4) Multidiscipline (n=4).

**Table 1**. Characteristics and distribution of the 79 included studies.

| **Characteristics** | **n** | **%** |
|---|---|---|
| SBDH **data source** | | |
|   Structured Data | 32 | 40.5 |
|   Unstructured Data | 23 | 29.1 |
|   Structured Data and Unstructured Data | 24 | 30.4 |
| **SBDH level** | | |
|   Individual | 49 | 62.0 |
|   Neighborhood | 7 | 8.9 |
|   Both | 23 | 29.1 |
| **Size of Data Sample** | | |
|   <10,000 | 36 | 45.6 |
|   Between 10,000 and 100,000 | 19 | 24.1 |
|   >100,000 | 15 | 18.9 |
|   Not specified | 10 | 12.6 |
| **Use of** SBDH | | |
|   Predictors for predictive modeling | 56 | 70.9 |
|   Disease management | 17 | 21.5 |
|   Outcome of analysis | 6 | 7.6 |
| **Publication Venues** | | |
|   Clinical | 31 | 39.2 |
|   Informatics | 33 | 41.8 |
|   Social science | 12 | 15.2 |
|   Multidiscipline | 4 | 5.1 |
| **Informatics methods to obtain SDOH** | | |
|   Predictive modeling | 23 | 29 |
|   NLP | 14 | 18 |
|   Statistical methods | 2 | 3 |
|   Use 2 methods | 28 | 35 |
|   Use all 3 methods | 13 | 16 |

SBDH**:** Social and Behavioral Determinants of Health**; NLP:** Natural Language Processing.

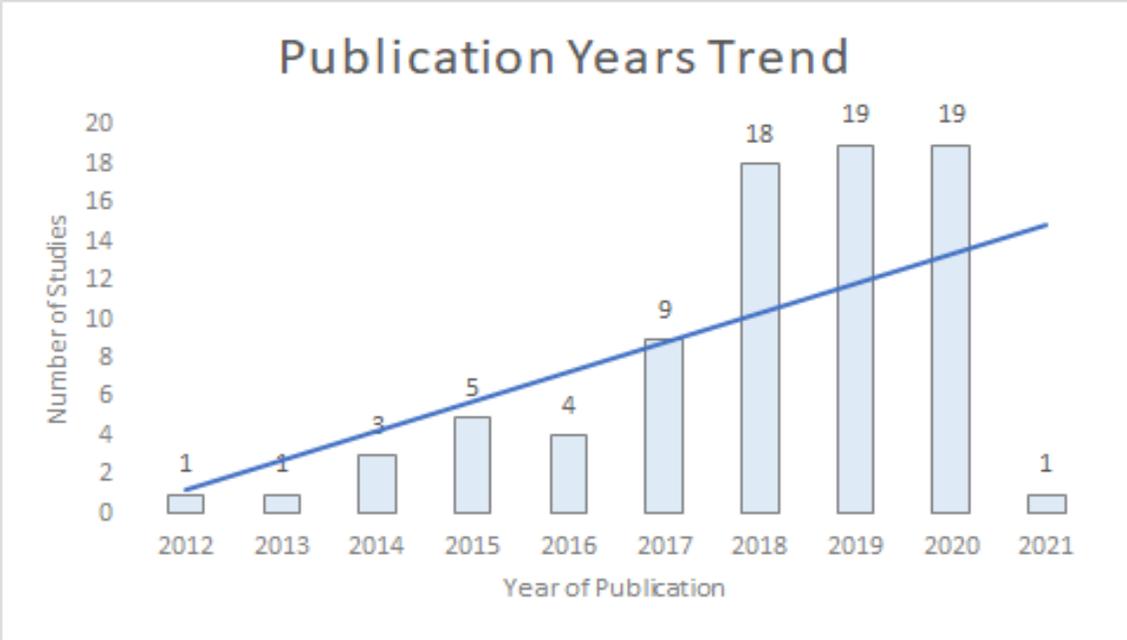

**Figure 2**. Years trend of 79 reviewed studies.

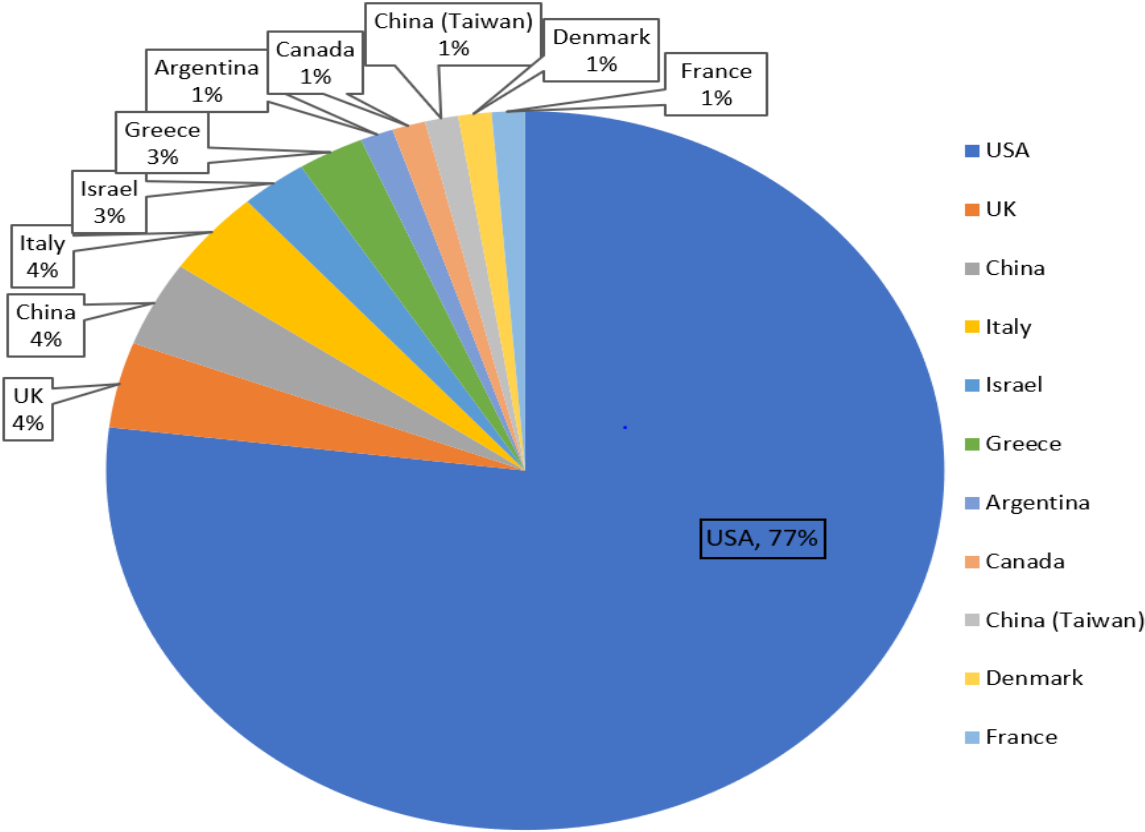

**Figure 3**. Summary of number of reviewed studies by country.

Electronic health records (EHR) are used in all studies in the review. Overall, 40.5% of the reviewed studies used structured data, 29.1% used unstructured data and 30.4% used both structured and unstructured data. Electronic health records (EHR) are used in almost all studies in the review, with other data from claims data[13]–[17], NHANES[17] and clinical trials[18]. Individual SBDH data were used in 62.0% of studies, neighborhood SBDH data were used in 8.9% of studies, and both individual and neighborhood data on SBDH were used in 29.1% of studies. Most of the studies (45.6%) have less than 10,000 data samples, 24.1% of studies have between 10,000 and 100,000 samples, and 18.9% have more than 100,000 samples.

Among all included studies, the most common usage (70.9%) of the SBDH information is acting as predictors for health outcomes, followed by disease management (21.5%) and outcome of analysis (7.6%). The percentage of nonclinical studies (48%) is slightly lower than that of clinical studies (52%). Natural language processing (NLP) and predictive modeling are the 2 main types of AI methods used to obtain SDOH information in the reviewed studies. 29% of the total reviewed studies used only predictive modeling, 18% used only NLP and 3% used only statistical analysis. The rest of the studies used either 2 of the methods mentioned above (35%) or all 3 methods (16%).

### 3.3 SBDH type

There is no single standardized method for categorizing SBDH factors. For example, the WHO[19] SDOH conceptual framework includes socio-economic and political context; socio-economic position; social cohesion and social capital; and health system. However, Healthy People 2030 [1] categorized SBDH factors into economic stability, education access and quality, social and community context, neighborhood and built environment, and healthcare access and quality instead. Since the articles we reviewed had no information on the socio-political context, we have classified the SBDH factors according to the Healthy People 2030 framework. Several studies have attempted to study SBDHs in order to determine the impact of social factors on health. The inclusion of SDOH in **Table 2** was determined by coverage of SBDH in one or more publications. From the articles reviewed, the SBDH factors identified were categorized into one of the five SBDH categories. While identifying SBDH factors, few SBDH mentions from the reviewed articles have been standardized to a single SBDH factor to prevent intense granularity of SBDH factors. For example, SBDH mentions such as alcohol, tobacco, and drug abuse have been normalized to substance use/abuse SBDH factor. Most of the studies focused on more than one SBDH factor belonging to different SBDH categories.

Overall, 29.5% of the studies reviewed focused on SBDH factors associated with healthcare access and quality, 24.7% focused on economic stability, 20% focused on social and community context, 16.3% focused on neighborhood and built environment, and 9.5% focused on education access and quality. Widely studied SDOH factors include substance use/abuse (9%) [14], [16], [20]–[35], education (7.3%) [16], [21], [36]–[46], employment status (6.3%) [16], [20], [21], [23], [31], [36], [38], [39], [45]–[48], socioeconomic status (6.3%) [29], [39], [44], [46], [49]–[56], lifestyle (5.8%) [22], [29], [32], [43], [57]–[63], socioeconomic factors (5.3%) [28], [38], [39], [46], [58], [63]–[67], diet (5.3%) [21], [22], [26], [43], [58], [68]–[72], housing status (5.3%) [16], [21], [33], [35], [38], [46], [73]–[76], social support (5.3%) [14], [15], [35], [74], [77]–[82], physical activity (4.8%) [20], [31], [40], [58], [68]–[70], [78], [83], marital status (3.7%) [31], [36], [40], [45], [55], [84], [85], housing instability (2.6%) [14], [16], [45], [75], [86], environmental factors (2.1%) [20], [52], [56], [87], insurance (2.1%) [20], [24], [38], [88], and homelessness (2.1%)[89]–[93]. Other significant SDOH factors include geographic location [24], [31], [44], [70], health literacy [43], [47], [88], [94], social and behavioral determinants of health [33], [52], [73], [95], social environment [41], [56], [77], health access [21], [54], [88], living condition [20], [31], [35], and social behavior [54], [63], [82], and financial insecurity [81], [86].

**Table 2.** The distribution of articles studying SBDH.

| SBDH Category | SBDH | No. of Articles |
|---|---|---|
| Economic Stability | Employment Status | 12 |
| | Socioeconomic Status (SES) | 12 |
| | Socioeconomic Factors | 10 |
| | Housing Instability | 5 |
| | Income | 2 |
| | Financial Insecurity | 2 |
| | Vocational History | 2 |
| | Work-related Challenges | 1 |
| | Housing Insecurity | 1 |
| Education Access and Quality | Education | 14 |
| | Health Literacy | 4 |
| Healthcare Access and Quality | Substance Use/Abuse | 17 |
| | Lifestyle | 11 |
| | Diet | 10 |
| | Physical Activity | 9 |
| | Insurance | 4 |
| | Health Access | 3 |
| | Erratic Healthcare | 1 |
| | Sleep | 1 |
| Neighborhood and Built Environment | Housing Status | 10 |
| | Homelessness | 4 |
| | Environmental Factors | 4 |
| | Geographic Location | 4 |
| | Living Condition | 3 |
| | Socio-environmental Neighbourhood | 2 |
| | Criminality | 1 |
| | Land Use Patterns | 1 |
| | Violence | 1 |
| | Weather | 1 |
| Social and Community Context | Social Support | 10 |
| | Marital Status | 7 |
| | Social and behavioral determinants of health | 4 |
| | Social Environment | 3 |
| | Social Behaviour | 3 |

|  | Psychosocial Factors | 2 |
|  | Social Isolation | 2 |
|  | Racial Disparities | 1 |
|  | Incarceration | 1 |
|  | Material and Social Deprivation | 1 |
|  | Social Activity | 1 |
|  | Social Characteristics | 1 |
|  | Social Circumstances | 1 |
|  | Social Discrimination | 1 |

## 3.4 Relations between SBDH and Health Outcomes

### 3.4.1 Healthcare related statuses

Among the 79 articles, 59 articles focused on SBDH factors associated with one or more disease conditions. From the 59 articles reviewed; the diseases investigated were grouped into one of the 12 categories (see **Table 3**). 25% of the studies focused on mental, behavioral and neurodevelopmental disorders, 17% on endocrine, nutritional and metabolic diseases, and 10% on diseases of the circulatory system. The most widely studied diseases include diabetes[45], [57], [59], [61], [66], [94], obesity[45], [55], [68], [72], geriatric syndrome[15], [58], [78], [80], and HIV[51], [82], [95]. Other notable mentions include hypertension[45], [60], stroke[45], [74], dementia[22], [40], and cancer[32], [96]. An interesting finding is that about 6 articles[23], [27]–[29], [35], [83] studied SBDH factors related to outcome measures such as hospital readmission risk, all-cause nonelective readmission.

**Table 3.** Number of articles for various healthcare related statuses.

| Category of healthcare related statuses | No. of Articles | Disease | No. of Articles |
|---|---|---|---|
| Certain infectious and parasitic diseases | 3 | HIV | 3 |
| Diseases of the circulatory system | 7 | Hypertension | 2 |
|  |  | Stroke | 2 |
|  |  | Acute myocardial infarction | 1 |
|  |  | Cardiovascular Disease | 1 |
|  |  | Congestive heart failure | 1 |
| Diseases of the digestive system | 1 | Crohn's Disease | 1 |
| Diseases of the musculoskeletal system and connective tissue | 1 | Gout | 1 |
| Diseases of the respiratory system | 4 | COVID-19 | 1 |
|  |  | Pediatric Asthma | 1 |
|  |  | Pneumonia | 1 |
|  |  | Seasonal influenza | 1 |
| Endocrine, nutritional and metabolic diseases | 12 | Diabetes | 6 |
| | | Obesity | 4 |
| | | Childhood obesity | 1 |
| | | Familial Hypercholesterolemia | 1 |
| Injury, poisoning and certain other consequences of external causes | 4 | Mild Traumatic Brain Injury (mTBI) | 1 |
| | | Non-fatal Suicide Attempt | 1 |
| | | Post-Deployment Stress | 1 |
| | | Suicide | 1 |
| Mental, Behavioral and Neurodevelopmental disorders | 17 | Dementia | 2 |
| | | Mental Health | 2 |
| | | AD Dementia | 1 |
| | | Alzheimer's disease | 1 |
| | | Attention-Deficit/Hyperactivity Disorder (ADHD) | 1 |
| | | Bipolar Disorder | 1 |
| | | Delirium | 1 |
| | | Depression | 1 |
| | | Epilepsy | 1 |
| | | Major Depressive Disorder | 1 |
| | | Mental Disorders | 1 |
| | | Opioid Misuse | 1 |
| | | Psychiatric Disorders that can lead to suicidal behaviors | 1 |
| | | Schizophrenia-Spectrum Disorders | 1 |
| | | Substance Use Disorders (SUDs) | 1 |
| Miscellaneous | 10 | Geriatric Syndrome | 4 |
| | | Multiple Diseases | 2 |
| | | Allograft survival | 1 |
| | | Chronic Diseases | 1 |
| | | Postnatal Growth | 1 |
| | | Total Knee Arthroplasty (TKA) | 1 |
| Neoplasms | 3 | Cancer | 2 |
| | | Prostate Cancer | 1 |
| Outcome Measures | 6 | Hospital readmission risk | 2 |
| | | 30-day all-cause nonelective readmission | 1 |
| | | 3-Day Postdischarge Readmissions | 1 |
| | | First emergency admission prediction | 1 |
| | | risk of 30 day readmission | 1 |
| Pregnancy, childbirth and the puerperium | 1 | Gestational diabetes mellitus (GDM) | 1 |

### 3.4.2 SBDH Categories

SBDH reflects social, physical, economic, environmental influences that can or cannot be regulated by the person but have a major effect on the wellbeing of the individual. The SBDH factors can act either as risk factors or intervention factors and can influence the burden of disease. There has been very little research on the association between SBDH characteristics and disease, as well as the prevalence of SBDH in any disease conditions. It is necessary to know the factors that influence the disease and to better understand the relationship, we have tried to draw attention to the diseases listed in the articles and their respective SBDH factors. **Figure 4** shows that the top 10 SBDH factors include education, substance use/abuse, socioeconomic status (SES), diet, lifestyle, social support, employment status, socioeconomic factors, marital status, and physical activity. Whereas the top 5 diseases include obesity, geriatric syndrome, diabetes, HIV and childhood obesity. The center part of **Figure 4** shows the associations between SBDH with a variety of diseases are shown in**.**

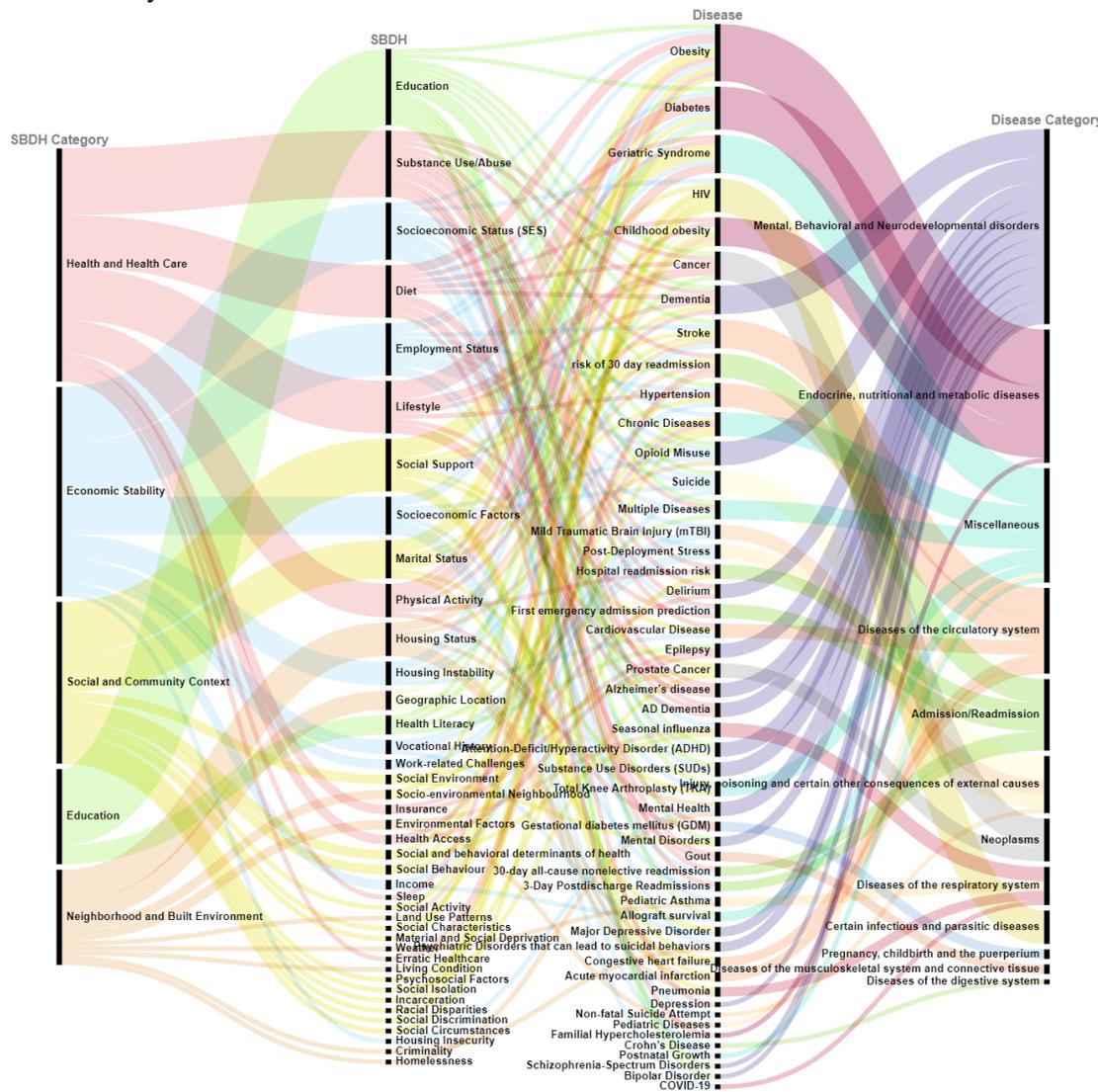

**Figure 4.** Overview of the relationship schema between SBDH and disease.

**Healthcare Access and Quality.** The SBDH factors such as substance use/abuse, diet, lifestyle and physical activity were mainly studied in endocrine, nutritional and metabolic diseases (diabetes and obesity) and mental, behavioral and neurodevelopmental disorders (Alzheimer's disease, dementia, mental disorders, delirium and depression). Two studies showed data integration through the semantic ETL service[68] and the MOSAIC dashboard system [57], using SDOH factors such as lifestyle, physical activity and diet could improve the management of obesity and diabetes. Hosomura et al. showed that educating patients with lifestyle interventions has been associated with improved glycemic control in diabetes patients [59], whereas Zhou et al. analyzed published lifestyle exposures and related intervention strategies for AD patients [22].

**Economic Stability.** The association between SBDH factors such as socioeconomic factors, employment status, SES and mental, behavioral and neurodevelopmental disorders (opioid misuse, ADHD, SUDs) and Injury, poisoning and certain other consequences of external causes (suicide, post-deployment stress) was widely studied. Zhang et al. and Afshar et al. studied the impact of low socioeconomic status and socioeconomic distress in individuals with at-risk comorbid SUDs [44] and opioid misuse [46] respectively. Zheng et al developed an early-warning system to identify patients at risk of suicide attempts[39] while a few studies stated that the predictors like SES[53], and socioeconomic factors [64] can be used to predict suicide risk.

**Education Access and Quality.** Education factor was critically analyzed in the mental, behavioral and neurodevelopmental disorders category (ADHD, bipolar disorder, dementia, mental disorders, opioid misuse, schizophrenia, substance use disorders (SUDs)). Education level in association with other factors like employment status, income found to have a significant correlation to suicidal behavior in patients with mental illness [39]. Senior et al developed OxMIS tool to predict suicide in patients with severe mental illness using SBDH factors such as highest education, substance abuse [42].

**Social and Community Context.** Although about 15 SBDH factors belong to this category, factors such as social support and marital status have been widely studied. Poor social support has shown to have an impact on hospital readmission [14], in patients with HIV [82] or Dementia [40]. Biro et al and Ge et al, respectively, studied the relationship between marital status and conditions such as obesity [55], suicidal ideation specific to major depressive disorder [85].

**Neighborhood and Built Environment.** Housing status including homelessness and geographic location were focused in the mental, behavioral and neurodevelopmental disorders category (delirium, ADHD, opioid misuse, substance use disorders (SUDs)) and diseases of the circulatory system (congestive health failure, acute myocardial infarction, stroke). Davoudi et al. and Nau et al. studied how geographic location serves as a surrogate of socioeconomic characteristics of the neighborhood, that have been shown to be associated with multiple diseases and health behaviors [24], and high mortality [70].

Among the individual diseases, obesity has been extensively studied, and the most researched factors include diet, marital status, education, employment status, housing instability, material and social deprivation, physical activity, sleep, SES and socio-environmental neighborhood. Social environment, education, financial insecurity, work-related challenges, lifestyle, and environmental factors were primary SDOH factors associated with mental disorders. Diabetes-related SDOH factors included lifestyle, socioeconomic factors, employment status, housing instability, education and marital status. Social support, physical activity, diet, lifestyle and socioeconomic factors were commonly studied with regard to geriatric syndrome. Most of the HIV-related SDOH factors fall within the social and community context, such as social behavior, social discrimination, incarceration, social support, and racial disparities.

### 3.4.3 The impact of SBDHs on disease

Among the individual diseases, obesity has been extensively studied, and the most researched factors include diet, marital status, education, employment status, housing instability, material and social deprivation, physical activity, sleep, SES and socio-environmental neighborhood. Biro et al. examined the association between obesity and material and social deprivation and found that patients in the most deprived group were 35% more likely to be obese than patients in the least deprived group[55]. Nau et al discovered 13 variables of the social, dietary, and physical activity environment that, when combined, correctly categorised 67 percent of communities as obesoprotective or obesogenic using mean BMI-z as a surrogate[70].

Social environment, education, financial insecurity, work-related challenges, lifestyle, and environmental factors were primary SBDH factors associated with mental disorders. Kim et al. reported that patients who abused alcohol were 3.3 times more likely to commit suicide than those who did not. According to adjusted predictive models, chart-noted alcohol abuse had a stronger association with suicide mortality than administrative data-based alcohol abuse diagnoses[30]. According to Davoudi et al., age, alcohol or drug misuse, and socioeconomic level can all increase the incidence of delirium[24]. According to Wang et al., the marital status of the dementia population was greater in the 'single' and 'widowed' categories, and a higher proportion in 'high school and equivalent' when compared to the non-dementia population[40]. Walsh et al incorporated risk factors such as comorbidities, medication use, clinical encounter histories, socioeconomic status, and demographics, and reported that machine learning models produced reliable prediction of non-fatal suicide attempts across several cohort comparisons and time frames[64]. Grinspan et al. demonstrated that bivariate analysis revealed multiple potential predictors of ED usage for epilepsy (demographics, social determinants of health, comorbidities, insurance, disease severity, and prior health care utilization)[88]. Zhou et al. demonstrated the viability of NLP techniques for the automated evaluation of a large number of lifestyle habits in patients with Alzheimer's disease using free-text EHR data[22]. Zhang-James et al. used machine learning algorithms to construct prediction models to identify ADHD youth at risk for SUDs by taking socioeconomic, educational, and geographic data into account[44].

Diabetes-related SBDH factors included lifestyle, socioeconomic factors, employment status, housing instability, education and marital status. Zhang et al. discovered new quantitative characteristics of electronic records of lifestyle counseling that are linked to better glucose control in diabetic patients[59].

Social support, physical activity, diet, lifestyle and socioeconomic factors were commonly studied with regard to geriatric syndrome. Anzaldi et al. discovered that the most common geriatric syndrome pattern among "frail" patients was a combination of walking difficulty, lack of social support, falls, and weight loss[78]. Kharrazi et al. used unstructured EHR notes enabled by the NLP algorithm to identify significantly higher prevalence of geriatric syndromes; 28 percent of patients lack social support[15]. Kuo et al. used machine learning to extract geriatric syndromes from EHR free-text in order to identify vulnerable older adults and possibly address functional and social inequities in the geriatric population[58].

Most of the HIV-related SBDH factors fall within the social and community context, such as social behavior, social discrimination, incarceration, social support, and racial disparities. According to Feller et al., socioeconomic determinants of health are increasingly acknowledged as predictors of HIV infection and were also included in the model by include words relating to drug use, housing instability, and psychological comorbidities. Structured EHR data, on the other hand, had a high variable relevance in the predictive models, and thus unstructured clinical text and structured EHR data exist as complementing sources of information for automated HIV risk assessment[95]. Wang et al. established the viability of using the EHR to quantify imprisonment exposure, a prevalent social determinant of health, for research purposes, particularly among racial and ethnic minorities and low socioeconomic status HIV patients[65].

### 3.5 NLP methods for extracting SBDH from clinical texts

To extract social determinants of health, there are various methods. For structured data, using database queries and descriptive statistics is a straightforward and frequently used method. For unstructured data, NLP is widely used to extract information. Information extraction is the task of automatically extracting predefined types of information, that is SBDH here, from unstructured or semi-structured data which is either EHR or clinical notes of patients' visiting a clinical site. In this review, 42 papers used NLP methods to extract SBDH from clinical notes.

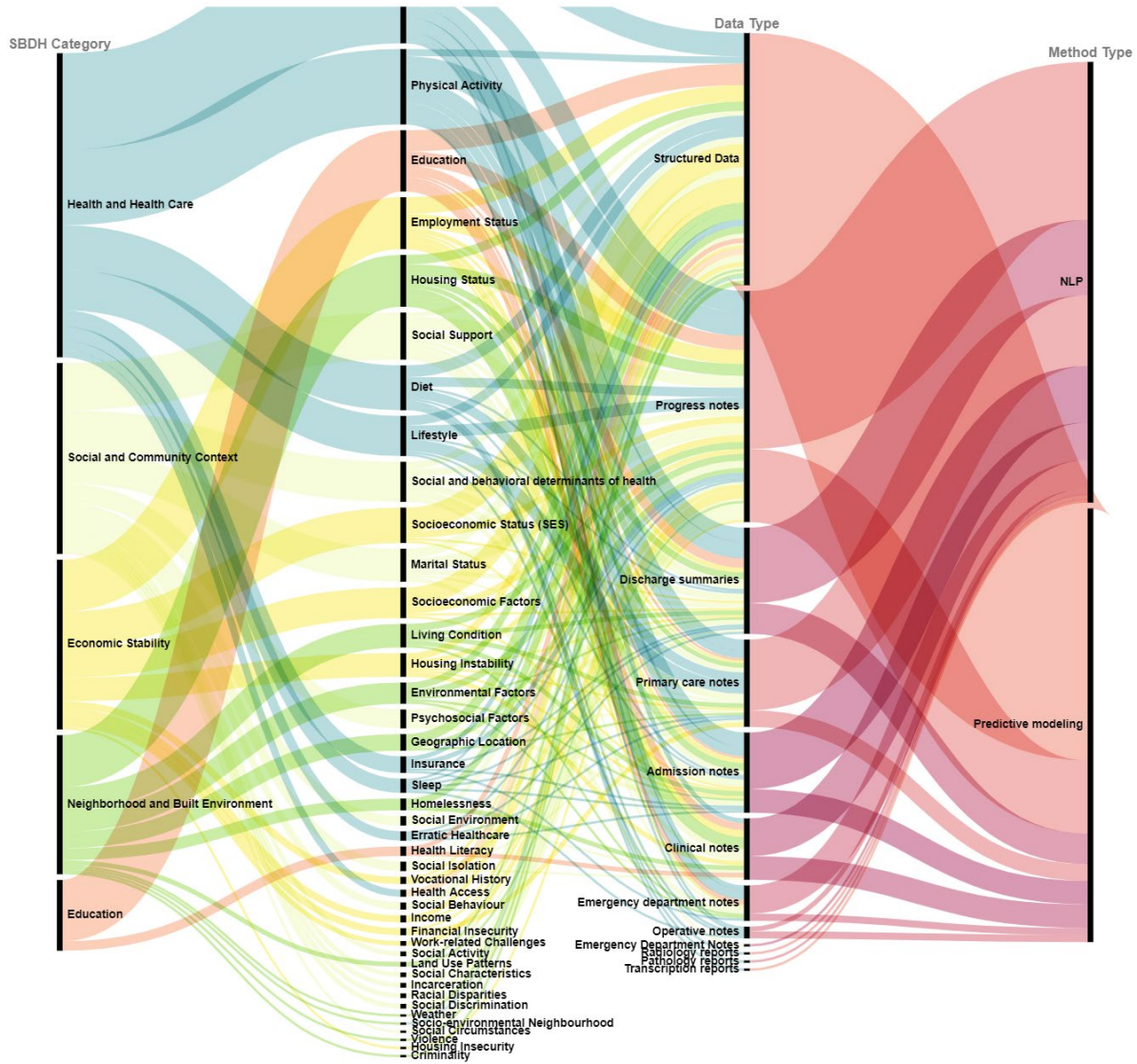

**Figure 5.** Overview of SBDH data source and AI methods.

**NLP Tools.** 42 articles used NLP tools including Apache cTAKES, MetaMap, Moonstone, MTERMS, Biomedicus, MediClass, and LEO[14]–[16], [20], [22], [26], [31]–[35], [35], [38], [40]–[42], [45], [46], [48], [51], [59], [61], [62], [68], [69], [71], [73], [74], [76]–[81], [83], [84], [86], [87], [90], [94]–[96]. Apache cTAKES was developed on the UIMA platform and Apache openNLP toolkit and is one of the most popular NLP tools for clinical information extraction from EHR data [97]. cTAKES was used to identify subtypes of patients with opioid misuse[46], and lifestyle modification [62]. Shoenbill et al. combines cTAKES with rules and regular expressions on selected EHRs to make previously unseen data on lifestyle modification documentation visible. They report that the results on testing the NLP tool refinement process for combined lifestyle modification retrieval were excellent with 99.27% recall and 94.44% precision and an F-measure of 96.79%[62]. MetaMap was originally developed to map biomedical literature to UMLS Metathesaurus concepts, but later applied to clinical texts. MetaMap was used for identification of homeless patients [16] and extraction of lifestyle information [22]. The Moonstone system is an open-source rule-based clinical NLP system designed to automatically extract information from clinical notes, especially those requiring inferencing from lower-level concepts[74]. The system was designed to extract social risk factors including housing situations, living alone, and social support [35], [74]. MTERMS [98] encodes clinical text using different terminologies and simultaneously establishes dynamic mappings between them. It was originally designed to extract medication information from clinical notes to facilitate real-time medication reconciliation, and later has been extended to support a variety of clinical applications, such as risk factor identification from physician notes [14]. The last study [14] used MTERMS to extract social factor information from physician notes and found that when compared to an 18.6 % baseline readmission rate, risk-adjusted analysis exhibited higher readmission risk for patients with housing instability (readmission rate 24.5 percent; p < .001), depression (20.6 percent; p < .001), drug abuse (20.2 percent; p = .01), and poor social support (20.0 percent; p = .01).

BioMedICUS[1] is an open-source system, built on the UIMA framework, for large-scale text analysis and processing of biomedical and clinical reports. It was used to process social history documents to identify social history topics [31]. Mediclass [99] is a knowledge-based system that can detect clinical events in both structured and unstructured EHR data. Hazlehurst et al. used the MediClass system with NLP components to identify the 5As of weight loss counseling [69]. The automated method [69] successfully recognized many valid cases of Assist that were not identified in the gold standard and on comparison with the gold standard, mean sensitivity and specificity for each of the 5As was at or above 85%, with the exception of sensitivity for Assist which was measured at 40% and 60% respectively for each of the two health systems. The Leo is a framework that was first built to support scalable deployment of NLP pipelines for processing a large amount of Veteran Affairs clinical notes. It facilitates the rapid creation and deployment of Apache UIMA-Asynchronous Scaleout annotators. The Leo system employed rules to identify the plan portion of the medical records and to detect words and phrases that capture instances of behavioral modification counseling [26]. Gundlapalli et al. demonstrates the feasibility of extracting positively asserted concepts related to homelessness from the free text of medical records with a two step approach. They first develop a lexicon of terms related to homelessness and then use the V3NLP Framework to detect instances of lexical terms and compare them to the human annotated reference standard. Their approach has a positive predictive value of 77% for extracting relevant concepts[90]. Schillinger et al. exploits NLP techniques to build scalable and reliable literacy profiles for identifying limited health literacy patients with respect to clinical diabetic patients' data. Using linguistic indices they developed two automated literacy profiles that have sufficient accuracy in classifying levels of either self-reported health literacy (C-statistics: 0.86 and 0.58, respectively) or expert-rated health literacy (C-statistics: 0.71 and 0.87, respectively) and were significantly associated with educational attainment, race/ethnicity, Consumer Assessment of Provider and Systems (CAHPS) scores, adherence, glycemia, comorbidities, and emergency department visits[94].

---

[1] https://github.com/nlpie/biomedicus3

**Rule-based methods.** Rule-based methods were widely used (7 studies)[15], [32], [38], [76], [78], [81], [83] for extracting SDOH information from clinical records. Rules including key terms were usually manually curated by domain experts. Hollister et al. used 860 terms to extract socio-economic status related SDOH data from free texts to demonstrate the feasibility of retrieving SDOH data and linking with other health related data for genetic studies [41]. A pattern-based NLP method was used to identify additional syndromes from clinical notes including malnutrition and lack of social support [15]. Incorporating unstructured EHR notes, enabled by applying the pattern-based NLP method, identified considerably higher rates of geriatric syndromes: absence of fecal control (2.1%, 2.3 times as much as structured claims and EHR data combined), decubitus ulcer (1.4%, 1.7 times as much), dementia (6.7%, 1.5 times as much), falls (23.6%, 3.2 times as much), malnutrition (2.5%, 18.0 times as much), lack of social support (29.8%, 455.9 times as much), urinary retention (4.2%, 3.9 times as much), vision impairment (6.2%, 7.4 times as much), weight loss (19.2%, 2.9 as much), and walking difficulty (36.34%, 3.4 as much). In [32], the researchers developed an information extraction system, which is based on a novel permutation-based pattern recognition method to extract information from unstructured clinical documents. The overall information extraction accuracy of the system for semi-structured text is recorded at 99%, while that for unstructured text is 97%. Furthermore, the automated, unstructured information extraction has reduced the average time spent on manual data entry by 75%, without compromising the accuracy of the system. In [38], they suggest some categories of socioeconomic status(SES) data are easier to extract in EHR than others. SES data extracted from manual review of 50 randomly selected records were compared to data produced by the algorithm, resulting in positive predictive values of 80.0% (education), 85.4% (occupation), 87.5% (unemployment), 63.6% (retirement), 23.1% (uninsured), 81.8% (Medicaid), and 33.3% (homelessness). Also, rule-based NLP methods can be used to extract information to represent patients in different groups. In [78], 2202 patients were described as "frail" in clinical notes. These patients were older (82.3 ± 6.8 vs 75.9 ± 5.9, $p< .001$) which had a significantly higher rate of healthcare utilization than the rest of the population($p <0.001$). In [81], the researchers used NLP to categorize reasons for social work referral documented in EHR referral orders. The most frequent needs leading to a social work referral were financial (25%), pregnancy (25%), behavioral health (16%), and family/social support (9%) needs. The most frequently co-occurring needs are pregnancy with language limitation (support = 0.07; confidence = 0.78); behavioral health with family/social support (support = 0.03; confidence = 0.28); and financial with behavioral health (support = 0.025; confidence = 0.14).

**Term expansion.** The limitations for rule-based methods is that they are hard to capture comprehensive lexical variation in the clinical records. Thus 4 studies investigated using unsupervised learning methods to further expand terms or get word representations from unannotated clinical texts. Word embeddings are such an approach, which is a type of word representations that allows semantically similar words to have a similar presentation based on the contexts of a corpus. Shi et al. used word2vec to retrieve the vector representation for each word in the EHR data, which was then added into a deep neural network model. Lexical association is another approach, a measure determining the strength of association between two terms in a text corpus [87]. Dorr et al. used lexical association approaches to expand psychosocial terms in clinical texts, helped them to identify a 90-fold increase in patients [86]. Bejan et al. implemented both lexical association and word2vec approaches to expand keywords of homelessness [73]. The word2vec was found to perform better (area under the precision-recall curve [AUPRC] of 0.94) than lexical associations (AUPRC = 0.83) for extracting homelessness-related words.

**Topic modeling.** Topic modeling is one of the unsupervised learning methods to explore latent topics in a given corpus. Three studies [40], [46], [95] used topic modeling for clinical notes. Latent Dirichlet Allocation (LDA) is a robust topic model, which learns K topics for a given corpus, where each topic is represented as a distribution of n words. Wang et al. used LDA to explore various themes (e.g., nutrition, social support) mentioned in care provider notes of dementia patients [40]. Among 250 topics generated by LDA models, they identified 224 stable topics. Some topics convey similar themes, and the domain experts analyzed all stable topics and classified them into 72 unique categories, such as medication delivery, and hospital care. Feller et al. used both LDA and TF-IDF to identify keywords for developing a prediction model on HIV risk assessment. These keywords are related to drug use and housing instability [95]. The patterns and trends of the generated topics provide unique findings and insights that are often not documented in the structured data fields in the EHR

**Deep learning.** A couple of studies developed deep learning methods [42], [80]. Chen et al. trained a deep neural network model using contextual information to identify sentences indicating the presence of a geriatric syndrome including lack of social support from clinical notes[80]. Contextual information improved classification, with the most effective context coming from the surrounding sentences. At sentence level, their best performing model achieved a micro-F1 of 0.605, significantly outperforming context-free baselines. At the patient level, their best model achieved a micro-F1 of 0.843. Senior et al. develop a neural network model to extract information such as the highest formal education from clinical notes as predictors for suicide in severe mental illness [42]. They developed a named entity recognition model, which recognizes concepts in free-text. The model identified eight concepts relevant for suicide risk assessment: medication (antidepressant/antipsychotic treatment), violence, education, self-harm, benefits receipt, drug/alcohol use disorder, suicide, and psychiatric admission. The named entity recognition model had an overall precision of 0.77, recall of 0.90 and F1 score of 0.83. The concept with the best precision and recall was medication (precision 0.84, recall 0.96) and the weakest were suicide (precision 0.37), and drug/alcohol use disorder (recall 0.61).

**Corpus development.** Developing a corpus is vital for developing reliable NLP methods. Three studies have focused on development of an annotated corpus on SDOH. Volij et al. developed an annotation standard to detect intra-social support from the electronic medical records [79]. Lybarger et al. recently used an active learning framework and developed the Social History Annotation Corpus (SHAC), including 4480 social history sections for 12 SBDH characterizing the status, extent, and temporal information [20]. The actively selected samples improve performance in both the surrogate task and the target event extraction task. A neural multi-task model is presented for characterizing substance use, employment, and living status across multiple dimensions, including status, extent, and temporal fields. The event extractor model achieves high performance on the MIMIC and UW Dataset:0.89-0.98 F1 in identifying distinct

SDOH events, 0.82-0.93 F1 for substance use status, 0.81-0.86 F1 for employment status, and 0.81-0.93 F1 for living status type.

**Conversational agent.** One study designed a study to analyze the significance of employing Alexa-based intelligent agents for patient coaching. Their study has shown that intelligent agents are another highly efficient model for intervention. Furthermore, they claimed that this approach has the potential to reshape the way people apply interventions [96].

## 3.6 Predictive models using SBDH for healthcare outcomes

Among these studies, 57 studies used SBDH factors for predicting healthcare outcomes. Predictive modeling is a technique that uses mathematical and computational methods to predict an event or outcome in a future time point of interest. In most cases a model is chosen based on a detection theory to try to guess the probability of an outcome given a set amount of input variables. In general, these models can either make use of one or more classifiers in order to determine the probability of a set of data belonging to another set or according to the undertaken task. Below we categorized these studies based on different methods of predictive modelling techniques. Note that one study can mention multiple predictive models and thus these studies were counted for each methodology category.

### 3.6.1. Supervised Learning Methods

**Regression Models.** Regression is the mostly used approach for predicting outcomes. Sixteen studies [21], [23], [25], [30], [36], [43], [45], [50], [52], [53], [56], [63], [65], [75], [91], [100] used various types of regression models, including logistic regression, LASSO regression, and Cox proportional models. Kim et al. used logistic regression to develop predictors for suicide using various suicidal behaviors and substance-related variables. After adjusting for administratively available data, they found that prescription drug misuse had an odds ratio (OR) of 6.8 (95% CI, 2.5-18.5); history of suicide attempts, 6.6 (95% CI, 1.7-26.4); and alcohol abuse/dependence, 3.3 (95% CI, 1.9-5.7) were major predictors for sucicide. Difficulty with access to health care was also a predictor of suicide (OR = 2.9; 95% CI, 1.3-6.3).

**Random Forest and Decision Tree.** Twelve studies [24], [28], [49], [54], [58], [60], [64], [70], [84], [92], [93], [95], used tree-based ML algorithms, including random forest and decision trees. Nau et al. utilized non-parametric machine learning methods such as Conditional Random Forest (CRF) to identify the combination of community features that are most important for the prediction of obesogenic and obesoprotective environments for children [70] whereas Agrawal et al. also used random forests along with gradient boosting methods and stacked generalization methods to attain their outcome using structured data [28]. Davoudi et al.[24], Walsh et al.[64],Grinspan et al.[88], Feller et al.[95], Erickson et al.[16], all make use of random forest variants. Grinspan et al.[88] showed through bivariate analyses that there are multiple potential predictors of emergency department use such as demographics, SDH, comorbidities, insurance, disease severity, and prior health care utilization. The paper used EHR data to predict ED use in two centers. Random Forest model with mean=2.9, and IQR=2.7-3.1 tied with a 3-variable model at one of the two centers and the latter out-performed at the remaining center.

**Neural Networks.** Eight studies utilized neural networks[27], [39], [44], [52], [53], [65], [87], [101]. Shi et al. [87] implements bidirectional RNN to predict pediatric diagnosis whereas Vrbaški et al.[101] develop predictors for lipid profile prediction. Both methods use a combination of structured and unstructured data with various natural language preprocessing steps. Xue et al. [45]utilized an RNN-based time-aware architecture to predict obesity status. An ensemble model with cross-sectional random forest (RF) model, a longitudinal recurrent neural network (RNN) model with the Long Short-Term Memory (LSTM) architecture is built in Zhang-James et al. [44]to predict at-risk comorbid SUDs in individuals with ADHD improvement. Subsequently, they found that population registry data and linked EHRs can be used reliably to predict at-risk comorbid SUDs in individuals with ADHD. Finally, one of the important empirical observations from this paper is that risk monitoring over years during child development can be achieved using a longitudinal LSTM model which was able to predict later SUD risks at as early as 2 years of age, 10 years before the earliest diagnosis with an average AUC of 0.63.

Support Vector Machines: Studies that used SVMs for prediction are relatively few (five studies)[26], [65], [72], [80] as compared to other methods which are discussed above. Wang et al. compares between back-propagation neural network (BPNN) , support vector machine (SVM), and logistic regression (LR) models to predict CD patients of non-adherence to azathioprine (AZA) and reports that SVM has the best performance[65]. Davoudi et al. uses various ML models including SVM to predict the risk of delirium using preoperative EHR data[24]. There are other interesting clinical studies like Wang et al.[65], which also used of SVM based methodologies.

### 3.6.2. Unsupervised Learning Methods.

A couple of studies used unsupervised machine learning methods. Afshar et al. [46]used LDA to identify subtypes of patients with opioid misuse whereas Cui et al. [67] used K-Means clustering and PCA to analyze and discover latent clusters in COVID-19 patients. Kirk et. al [61] discusses an algorithm using unsupervised markov clustering (MCL) and performs a phenotypic characterization of a Danish diabetes cohort. The stratification of the diabetes cohort is based on characteristics extracted from the unstructured EHR records of the target (homogenous) population, where these characteristics include several diagnoses and lifestyle factors. Patient clusters are obtained by exploiting unsupervised MCL along with other NLP techniques.

Thus, predictive modeling techniques are used to develop markers for predicting required clinical outcomes using a set of characteristics and behaviors. In most of the studies various SBDH are either part of features or characteristics that define a specific cohort and act as markers of a pre-defined outcome. There are also some studies that analyze the effect of adding SBDH into risk prediction models and whether it improves prediction accuracy for some specific ailments. In another line of research they analyze individual contributions of SBDH at the patient level, informing appropriate interventions that can reduce the risk of negative health outcomes such as preventable readmissions and/or hospitalizations.

## 4 Discussion

SBDH research has become an active and interdisciplinary research domain, covering healthcare, informatics, computers science, and social science. It is important to acknowledge that SBDH factors have a major impact on health outcomes. Our review indicates that various SBDH categories have been investigated with a wide range of disease categories. The most studied SBDH factors are substance abuse, employment status and socioeconomic status, whereas other important SBDH factors are under-studied, such as social environment, psychosocial factors, racial disparities et cetera potentially due to the lack of data. The most studied disease areas using SBDH are mental and endocrine disorders. Other severe diseases (e.g., cancers) for which SBDH could be important factors for healthcare outcomes are rarely studied using the extensive information in EHRs. The associations between socioeconomic factors and health outcomes are complicated and diverse; several pathways may be involved [102]. We observed that multiple SBDH factors are being investigated in a single disease and among the many factors, the number of SBDH factors that may potentially affect the condition of the disease is still questionable. The wide range of SBDH factors to be considered while examining the patient could indeed overwhelm the physician and may have an effect on decision-making and policy-making[103]. From our analysis, we found that most of the studies we reviewed focused on mid-and downstream SBDH factors and not upstream, like governance and policy [104]. Research on how SBDH influences establish pathways that contribute to health inequalities is much required.

EHR has rich information on patients' health conditions and treatment process; however, the representation of SBDH in EHR is still limited. Certain data is clearly stated in clinical text (for example, documentation of drug and alcohol use); nonetheless, a considerable proportion of information about certain SBDH factors such as social environment is not directly indicated in the clinical note but can be inferred. It is anticipated that the majority of clinical data in the EHR is unstructured and hence difficult to examine. Data such as physician notes, nurse notes, discharge summaries, and patient-reported information have the potential to contribute a plethora of essential clinical information, but are often unusable, depriving a significant method of improving population health. Using Natural Language Processing (NLP) technology to support the extraction of SBDH and other clinical ideas simplifies the identification and extraction of essential concepts from clinical data, efficiently unlocks unstructured data, and aids in the resolution of unstructured data-related issues. Better-informed population health decisions can be made with the use of accurate and complete SBDH information.

In this review, most studies focus on one or more SBDH factors. We can observe from the content related to the association between SBDH and healthcare related statuses that SBDH is recognized to have a potential effect to comprehend patients' health status. Even with several papers focusing on development of NLP algorithms to extract SBDH from clinical notes, there is still a data bias regarding the representation and completeness of SBDH in EHR. Individual level SBDH information usually contributes to the accuracy of predictive models; however, they are usually hard to capture accurately or completely in EHR. In this review, we only focused on NLP techniques that were used to extract one or more SDOH aspects from clinical records. Our analysis indicates that information extraction for SDOH has been dominated by rule-based approaches, including rule-based NLP tools and methods. Half of the studies used existing clinical NLP tools, several of which (e.g., cTAKES, MetaMap) were widely used in other domains as well. Manually curated key terms were also widely used for rule-based methods. These findings are consistent with the recent review on NLP methodology for clinical IE [97]. Due to the variability of clinical concepts and limitation of hand crafted term list, unsupervised machine learning methods (word embeddings) were commonly used to expand term lists by finding their semantically similar terms in the clinical corpora. Topic modeling was another commonly used approach to cluster key terms in a coherent and latent topic. These methods often need manual checks to confirm the final term list or appropriate number of topics. Very few clinical data corpora on SDOH were available, which leads to limited studies using supervised machine learning methods. However, there were a couple of studies investigating how to develop SBDH-specific corpus. One study utilized active learning methods, which has demonstrated to save human effort for annotations in other studies [105]–[107]. Deep learning models have shown promising results and predominate in the general NLP domain; however, only 2 studies were found to use deep learning methods in the analyzed literature. Developing accurate deep learning models in SDOH requires a large amount of annotated training data, which is a time-consuming and labor intensive process. One possible solution is to use advanced IE techniques, such as distant supervision [108], which automatically or semi-automatically generate weak labels for training deep learning models. Predictive models were widely used to investigate the association between SBDH factors and health outcomes. Such outcomes include specific diseases and administration aspects, such as readmission. There are national and regional efforts contributing to the integration of SBDH into EHRs, including establishing national standards[109] for SBDH data collection and representation [109], [110] and developing SBDH integration tools [11]. Various SBDH integration tools have emerged in order to collect more SBDH documentation in EHR. Thus, there are still a lot of efforts the community should work together to address SBDH data bias in EHR data.

AI has impacted every aspect of our daily life, from product recommendations to intelligent personal assistants, powered by the availability of large volumes of data. The increasing adoption of EHR systems in healthcare organizations has fostered the secondary use of EHR data in AI techniques to improve patient care outcomes, through clinical decision support, chronic disease management, patient education, and so forth. However, similar to human beings, AI algorithms are vulnerable to biases that may result in unfair decisions. For example, the Correctional Offender Management Profiling for Alternative Sanctions (COMPAS) system produces a higher risk score to African-American compared to Caucasians when it is used by judges to measure the risk of committing another crime[111]. In the context of healthcare settings, bias in AI algorithms may result in more serious issues, such as affecting patients' safety, delaying treatment, even risking lives. Therefore, addressing bias in AI algorithms is crucial for successfully deploying AI applications in healthcare and clinical practice [112]. AI bias can be summarized into two categories, i.e., data bias and algorithm bias. While algorithm bias is related to the algorithm design and model architecture that is not trivial to mitigate, data bias is relatively easier to address by carefully selecting unbiased cohorts and datasets. SBDH could serve as a metric to measure the bias in EHR data and select a fairness cohort to train AI algorithms. For example, socioeconomic status, an important SBDH variable, could be used to ensure training data better represents patients with different socioeconomic status.

Health equity issues caused by complex, integrated, and overlapping social structures and economic systems are gaining recognition among scientists and public health professionals. SBDH is an important indicator for health equity since it indicates whether people have access to adequate diet, medical care, educational and career opportunities, what are their healthy environmental conditions, and whether a person is exposed to physical or psychologic trauma [113]. SDOH helps us develop comprehensive strategies to address potential risks for the population, particularly for the under-served populations. It is well documented that the social conditions impact premature mortality in under-served communities. SBDH are responsible for many of the leading health disparities in the US. Gaining insight on the SBDH and how SBDH information could be extracted from EHRs improves the opportunities to increase wellness, prevent premature illness, gives health care teams the insight needed to increase patient action (i.e., adherence, behavior change, and compliance), provides needed information to influence health policy change for wellness, and eventually promotes health and health equity [6].

This review has several limitations. First, our search terms and databases might be insufficient to cover all studies. We only included articles written in English. Due to the fact that the definition of SBDH is very broad and not specifically defined in the literature, the searching keywords used in this review might not be sufficient for searching all SBDH-related articles. Second, we only focus on studies that developed or adapted AI methods in EHR data for SBDH research. Several studies utilizing non-EHR data to study SDOH, such as clinical trial data [18], survey data [114] or claims data [13][17], were excluded from this review.

# 5 Conclusion

In summary, this scoping review discussed the current trends, challenges, and future directions on using AI methods on SBDH in the EHR data. Our analysis indicates that despite known associations between SBDH and disease, SBDH factors are not commonly examined as interventions to improve the patient's healthcare outcomes. Gaining insights into SBDH and how SBDH data could be extracted from EHRs using NLP techniques and predictive models improves the opportunities to influence health policy change for patient wellness, and eventually to promote health and health equity.

# 6 Acknowledgement

YW was supported by the Mayo Clinic Center for Health Equity and Community Engagement Research Award. RZ was partially supported by the National Institutions of Health's National Center for Complementary & Integrative Health (NCCIH), the Office of Dietary Supplements (ODS) and National Institute on Aging (NIA) grant number R01AT009457 (PI: Zhang).